\documentclass[reprint, amsmath,amssymb,aps]{revtex4-1}
\usepackage{graphicx}
\usepackage{dcolumn}
\usepackage{bm}
\usepackage{dcolumn} 
\usepackage{bm} 
\usepackage{amsmath}
\usepackage{amsfonts} 
\usepackage{amssymb}
\usepackage[utf8]{inputenc}
\usepackage{textcomp}
\usepackage[english]{babel}
\usepackage[usenames,dvipsnames]{xcolor}
\usepackage{courier}
\usepackage{etoolbox}
\usepackage[normalem]{ulem}
\usepackage{listings}
\usepackage{color}

\begin{document}
\title{Harnessing the Power of the Second Quantum Revolution}

\author{Ivan H. Deutsch}
 \affiliation{Center for Quantum Information and Control. \\ Department of Physics and Astronomy. \\ University of New Mexico }
\date{\today}

\begin{abstract}
The second quantum revolution has been built on a foundation of fundamental research at the intersection of physics and information science, giving rise to the discipline we now call Quantum Information Science (QIS).  The quest for new knowledge and understanding drove the development of new experimental tools and rigorous theory, which defined the roadmap for second-wave quantum technologies, including quantum computers, quantum-enhanced sensors, and communication systems.  As technology has matured, the race to develop and commercialize near-term applications has accelerated. In the current regime of Noisy Intermediate Scale Quantum (NISQ) devices~\cite{Preskill18}, the continued necessity of basic research is manifest.  Under what conditions can we truly harness quantum complexity and what are its implications for potential useful applications?  These questions remain largely unanswered, and as the QIS industry ramps up, a continuous feedback between basic science and technology is essential.  In this Perspective I review how curiosity-driven research led to radical new technologies and why the quest for basic understanding is essential for further progress.  

\end{abstract}

\maketitle

\section{Introduction}
The end of the twentieth century saw the convergence of two of its major intellectual achievements: quantum mechanics and information science. This marriage gave birth to the field of Quantum Information Science (QIS) which has ignited a ``second quantum revolution" that promises next-generation information processing technologies that can far outperform current systems based on technologies that arose in the ``first quantum revolution," e.g., semiconductors and lasers.   Today, the acceleration of the second quantum revolution is palpable.  In the United States, the passage of the National Quantum Initiative (NQI) Act of 2018~\cite{NQI18} codifies a call to arms, with the goal of expanding the number of students, educators, researchers, and practitioners with training in QIS, with a particular eye towards increasing the quantum workforce. Similar programs have been established worldwide, including the UK National Quantum Information Programme, the EU Flagship on Quantum Technologies, and the Chinese National Laboratory for Quantum Information Sciences.  A phase transition has occurred in the private sector with rapid new investments in the development of quantum technology at all levels of industry, ranging from tech giants, to defense contractors, to startups.  The competition for talent is heating up, with academics and industry now battling to attract QIS-trained personnel.  The revolutionaries are marching!

Given this rapid acceleration, it is useful to revisit how we got here, and what will be necessary to fulfill the promises of the revolution.  QIS arose from curiosity-driven research which established the foundation of the field.  Questions at the foundations of quantum mechanics, which had been put to rest by most physicists in the 1930s, were revisited starting primarily in the 1960s, and put to the test in the 70s, 80s, and 90s.  Is quantum randomness explained by local hidden variables that encode elements of reality~\cite{EPR35, Bell64}?  Why are superposition and interference of macroscopic states not observed~\cite{Zurek03}?  Do quantum jumps occur~\cite{Schrodinger53, Nagourney86, Sauter86, Bergquist86, Gisin92, Minev19}? How does a quantum measurement occur in continuous time~\cite{Gisin92a, Carmichael93,Minev19}? While the formalism of quantum mechanics was well established by the 1920s, its full implications were not understood.  This new understanding grew out of a series of new (and/or re\"{e}xamined) concepts including entanglement~\cite{EPR35, Schrodinger35, Bell64, Freedman72,Aspect81,Horodecki09}, contextuality~\cite{Kochen67,Mermin93, Raussendorf2013}, negativity of quasiprobabilities~\cite{Kenfack04,Ferrie10,Veitch12}, decoherence~\cite{Zurek03}, no-cloning~\cite{Wootters82}, and quantum trajectories~\cite{Dalibard92, Dum92, Gisin92a, Carmichael93}.  

Putting these concepts to the test in the laboratory was a new form of fundamental research, different from the search for new states of condensed matter or new physics beyond the standard model.  The ability to control individual quanta of fields and matter~\cite{Haroche13, Wineland13} required new ground-breaking experimental tools such as laser cooling and trapping~\cite{Phillips98}, nonclassical light sources such as squeezed light~\cite{Walls83} and entangled photons~\cite{Shih03}, cavity QED~\cite{Miller05,Walther06,Haroche06}/circuit QED~\cite{Blais20}.  While such experiments did not cause us to rewrite the rules of quantum mechanics, they did create a new framework for thinking about them.  What is the power of nonlocality in quantum mechanics~\cite{Eberhard89, Popescu14}? In what sense is the quantum state a state of knowledge, and how should we update the state conditioned on the knowledge~\cite{Fuchs13}?  How is quantum coherence lost when going from the microscopic to the macroscopic~\cite{Brune96, Zurek03}?  Such a new framework is important because it opens our minds as to what is possible in principle, and how we might achieve it.  Plus, experimental techniques were now in place to tackle new challenges.

The other pillar of QIS, information science, also stands on a foundation of fundamental research.  Turing introduced Turing machines in the context of research into the foundations of mathematics in order to prove the uncomputability of certain functions~\cite{Turing36}.  Together with Church~\cite{Church36}, they abstracted computers from the machines themselves, to state that all reasonable models of computation were essentially equivalent and could be mapped to a universal Turing machine (the Church-Turing thesis).  Shannon took this to an even more radical extreme.  In his revolutionary work, he abstracted all of information from the machines that process it, with the invention of the bit, and a definition of information in terms of the observer and their ``surprise" from what they learn~\cite{Shannon48}.  Information is subjective!  The foundation of information theory was thus divorced from physics and machines.  But the role of physics in the foundation of information science was lurking in the background, starting with thermodynamics.  The mere fact that Boltzmann's entropy and Shannon entropy contain the same mathematical formula $-\sum_i p_i \log p_i$ (modulo Boltzmann's constant) gives strong indication of the intimate relationship between the two.  This was made rigorous by Jaynes, who showed how statistical mechanics should be considered as a theory of Bayesian statistical inference rather than a theory of physical law~\cite{Jaynes57}.  

Physics reasserted itself into information theory with the resolution of the Maxwell Demon Paradox.  The Szilard engine demonstrated that the process through which the Demon extracted information would have thermodynamic consequences~\cite{Szilard29}.  The complete resolution was put forward by Bennett~\cite{Bennett82} through Landauer's principle~\cite{Landauer61}. Unlike Szilard's engine, entropy need not be increased through the Demon's method of gaining information, which could be done reversibly, but it will be increased when the information is erased on the Demon's finite tape.  Information processing does indeed depend on the physics of the device that processes the information.  In this sense, information is physical!  

The thermodynamics of computation thus played a central role in the foundations of QIS. As shown by Bennett and others, computation itself could, in principle, be done reversibly, with no thermodynamic cost.  This led to the first suggestion by Benioff that a closed, reversible quantum system could perform a computation~\cite{Benioff82}.  Benioff's computer, however, simulated a (reversible) classical Turing machine.  Soon after Feynman~\cite{Feynman82} and Deutsch~\cite{Deutsch85} (no relation)  realized that a fully quantum Turing machine could have power beyond a Turing machine obeying classical physics.  The potential of a quantum computer to solve problems more efficiently than a Turing machine would rock the foundation of the Extended Church-Turing thesis~\cite{Yao03}.

The potential promise of quantum computing represents one amongst many of the promises of the second quantum revolution.   Quantum mechanics was often characterized as a paler version of classical mechanics, due to its intrinsic uncertainty and stochasticity, in contrast with the clockwork precision and determinism of Newtonian trajectories.  Quantum mechanics was viewed as a nagging parent always telling you what you can't do.  You can't know a particle's position and momentum at the same time.  You can't measure a system without disturbing its state.  The foundational work of the latter twentieth century showed that this negative characterization could not be further from the truth. Quantum mechanics is an enabler.  The inability to clone a quantum state~\cite{Wootters82} can make encryption unconditionally secure, be it for bank notes~\cite{Wiesner83} or quantum key distribution for communication channels~\cite{Bennett84}.  Quantum nonlocality allows for teleportation of information between distant parties~\cite{Bennett93}.  The quantum version of Fisher information dictates the ultimate noise floor for parameter estimation~\cite{Braunstein94} at the heart of precision sensing and metrology; nonclassical probes allow us to beat the standard limits imposed by uncorrelated corpuscular quanta~\cite{Giovannetti06, Pezze18}.  And quantum interference between an exponentially large number of outcomes in a quantum register can enable algorithms that we expect can solve problems much more efficiently than those believed to be solvable on a Turing machine, e.g., Shor's algorithm for integer factoring~\cite{Shor94}. 

The story of the birth of QIS is one of the greatest success stories of interdisciplinary science.  The foundation of physics and information theory are fundamentally connected.  Today, we examine whether quantum physics itself should be understood as essentially a theory of Bayesian statistical inference~\cite{Fuchs13}, in the same way that Jaynes considered classical statistical physics~\cite{Jaynes57}.   At the same time, information cannot be completely abstracted from physics.  The ability to process information, be it to estimate parameters (sensing/metrology), communicate/hide messages, or compute functions is intrinsically tied to the physics of the device that does the job.  Quantum physics enables a broader class of devices that potentially have more power than classical information processors.  

But, how do we harness this power?  While we believe that quantum mechanics governs all of physics, only a subset of devices can access its full potential.  Indeed, from the earliest days, skeptics rightfully questioned whether a quantum computer could supersede the power of a classical Turing machine, even in principle, as such a device was essentially analog and would be exponentially sensitive to a continuum of control errors~\cite{Landauer96}, quantum chaos in many-body dynamics~\cite{Georgeot00}, and critically, exponentially fast decoherence due to interaction with the environment~\cite{Unruh95, Chuang95}.  These challenges led Haroche and Raymond to famously muse whether quantum computing was a dream for theorist but a nightmare for experimentalists, because exponential speed up would certainly be associated with exponential complexity in implementation~\cite{Haroche96}.  

The dream is still alive.  The fact that, in principle, quantum coherence could be maintained for an arbitrarily long time, and with only a polynomial resource overhead is one of the most profound results of QIS, as first shown by Shor~\cite{Shor95}.  It was a huge surprise to twentieth century physicists that a system can be engineered to protect the wave function from ``collapsing" in the presence of a noisy environment, and that quantum interference could be extended into the macroscopic world.  This result took interdisciplinarity and a new perspective, which revolutionized our understanding of quantum mechanics.  The rigorous theory of quantum error correction and fault-tolerance sets the goalposts for universal quantum computing~\cite{Shor96, Knill96, Aharonov96, Gottesman97}.  The challenges are substantial but in principle possible, and there is great progress being made every day in significant hardware improvements, advanced quantum codes, and techniques for fault-tolerance such as flag qubits~\cite{Chao18}, bosonic encoding~\cite{Joshi20}, and single-shot error correction~\cite{Bombin15}.  Experiments in the Schoelkopf and Devoret groups with superconducting qubits coupled to microwave cavities have made first steps in demonstrations of true error correction, which reduce the rate of error below the native error rate~\cite{Ofek16, Campagne20}, and the first fault-tolerant operation on a logical qubit was demonstrated in the Monroe and Brown groups with trapped atomic ions~\cite{Egan20}.

Today we live in the era of Noisy Intermediate Scale Quantum (NISQ) information processors~\cite{Preskill18}. Substantial improvements in experimental techniques and hardware have enabled researchers and developers to scale up systems with 10-100 qubits to have relatively high-fidelity operations.  What lies at this ``entanglement frontier"~\cite{Preskill12}?  The development of near-term quantum technologies based on NISQ systems is one of the hottest topics today, with potential applications including quantum simulators of many-body physics~\cite{Feynman82, Lloyd96, Lewenstein07, Bloch12, Blatt12, Schaetz13, Georgescu14, Johnson14, Bernien17, Monroe19}, variational quantum eigensolvers for quantum chemistry and electronic structure~\cite{McClean16}, quantum optimizers~\cite{Albash18, Moll17, Brando16}, and quantum machine learning~\cite{Biamonte17}.  However, because of their intermediate scale, these systems are essentially analog machines, and we cannot employ the full machinery of fault tolerance error correction to digitize quantum information and protect it from inevitable noise.  Thus, to deliver on these technologies, continued basic research is essential.  How much quantum complexity can we truly generate with a NISQ device, and what are the conditions that must be reached to harness this complexity for a true quantum advantage? To answer these questions a new round of curiosity-driven research is in order.  In this perspective I will share further examples about how our drive to understand the fundamentals led to the foundations of technology, and where some of the mysteries remain that demand our curiosity.

\section{NISQ computing: dream or nightmare?}
\subsection{Classical Computing:  Digital vs. Analog}

The power of computation depends on the physics of the machine that carries out the computation.  This is QIS's challenge to the Extended Church-Turing thesis~\cite{Yao03}.  In thinking about models of computation, we typically divide them into two classes, analog and digital.  The distinction between these two models is not always clear cut, and various hybrid models  straddle these categories.  The situation becomes even murkier in the context of quantum computing, as we will see below. For the purpose of discussion here, I will make a strict division according to the following  definition:

\vspace{5pt}
\noindent
\textbf{Definition}: A model of computation is categorized as {\em digital} if the state space of acceptable inputs and outputs is countable (discrete) and the set of allowed functions on the input can implemented via a discrete set of steps (transitions functions), where all possible functions are specified by a discrete set of parameters.  Otherwise it is {\em analog.}
\vspace{5pt}

\noindent
This definition requires both the digitization of the state space and the specification of the dynamical map that performs the computations.  It also characterizes all hybrid models as analog. 

Traditional classical analog models of computation describe machines that find solutions to differential equations in continuous time.  An example is the General Purpose Analog Computer (GPAC) constructed by Shannon to describe mechanical and electrical differential analyzers as programmable computers consisting of set of basic components: adders, multipliers, integrators, amplifiers, etc~\cite{Shannon41}.  Such models were called ``analog" in that one physical system was the analog of the other; the same physical equations of motion apply in both systems, and the solutions to these equations are measured in the analog device, thus carrying out the computation.  By the definition above such a model is an analog computation for multiple reasons.  The state space is the real numbers,  the transition function from input to output is specified in continuous time, and there is a continuum of possible components, each specified by real-number parameters.  Hybrid models, such as recurrent neural networks~\cite{Siegelmann94} and the Blum-Shub-Smale model~\cite{Blum89} (a RAM machine with sequential gates in which a register can store an arbitrary real number with infinite precision) are analog machines; they are discrete in time but involve a continuous state space and a continuous set of dynamical maps.  

The Turing machine is the quintessential digital model of a classical computer~\cite{Turing36}, equivalently realizable in the von Neumann architecture through a universal logic gate model for implementing Boolean functions over a state space of $n$ bits~\cite{Zargham96}. Every digital model is an abstract model; all physical computers have a fundamentally analog nature.  The physical input-output map is implemented in a dynamical process that occurs in continuous time. And, in the context of classical physics, the phase space of possible states of particle and field degrees of freedom is continuous.  Digitization is implemented by {\em coarse-graining}, discretizing the state space, e.g., thresholding the voltage on a transistor in order to distinguish 0 and 1 as a two states of bit. Logic gates are obtained by coarse-graining the possible continuous analog evolutions into a discrete finite set.

While all digital computers that are governed by classical physics are considered to be equivalent according to the Extended Church-Turing Thesis~\cite{Yao03}, the case of analog classical computers is less clear cut.  Various analog models can exhibit super-Turing decidability~\cite{Copeland02, Bournez06}.  Moreover, analog models can have different complexity~\cite{BenHur02, Bournez16}.  Vergis {\em et al.} showed that a model of computation implemented with mechanical gears could solve 3-SAT, an NP-complete problem~\cite{Vergis86}.  Similarly, the problem of finding Steiner trees in graph theory is NP-hard, but can be extracted from the minimum energy configuration of soap bubbles in a certain geometry~\cite{Garey76}.  Such computational power derives from unphysical resources built into the analog model, as we will see below.  As Vergis {\em et al.} conclude, ``if a strongly NP-complete problem can be solved by an analog computer, and if $P \neq NP$, and if the Strong Church’s Thesis is true, then the analog computer cannot operate successfully with polynomial resources"~\cite{Vergis86}. Similarly, minimization of energy configurations will generally take exponential time in the size of system.  As Aaronson suggests, ``[$P \neq NP$, could] eventually attain the same status as (say) the Second Law of Thermodynamics, or the impossibility of superluminal signaling. In other words, while experiment will always be the last appeal, the presumed intractability of NP-complete problems might be taken as a useful constraint in the search for new physical theories"~\cite{Aaronson05}.  This conjecture implies that models of analog classical computing that allow efficient solutions to NP-hard problems must have a hidden physical cost. 

An ideal analog computer has intrinsic cost because the inputs, outputs, and/or dynamical maps are specified in the continuum.  The computational power of such analog devices is thus inextricably tied to the question of robustness.  All aspects of physical operation will be subject to imperfection:  state preparation (input), dynamics (transition function), measurement (output).  Physical sources of error include finite temperature, imperfect calibration, environmental background noise, and finite signal-to-noise in measurement. All of these imperfections lead to uncertainty and thus finite resolution to which we can assign an ideal state and dynamics of the analog machine. A key issue, thus, for the reliability of the analog computing is how uncertainty propagates.  This is particularly critical for solutions to complex problems, where dynamics can be nonlinear, potentially chaotic, and thus hypersensitive to noise. Digitization provides intrinsic robustness through coarse-graining. The state space and dynamical maps are discretized and thus accommodate finite resolution. Small noise and errors do not accumulate. When strong noise occurs, outside the typical range tolerated by the coarse graining, error correction techniques can be employed to recover ideal operation, without the loss of information or hypersensitive noisy dynamics.

Are experimental imperfections and concomitant uncertainty just a nuisance that can be solely attributed to human error or does something more fundamental lurk here?  As von Neumann argued in his analysis of the effect of noise on automata, ``error should be treated by thermodynamical methods, and be the subject of a thermodynamical theory, as information has been, by the work of L. Szilard and C. E. Shannon''~\cite{vonNeuman56}.  I posit that the resources necessary to achieve a required degree of fine graining (precision) is an intrinsic resource for reliable computation, similar to Landauer's principle~\cite{Landauer61}.  The degree of precision required will depend on the model of computation (analog vs. digital), with some models robust, and others hypersensitive.  The hypersensitive models could require exponentially large resources, and thus render their operation inefficient for some applications.

\subsection{Quantum Computing: Digital vs. Analog}

Are quantum computers digital or analog according to the definition above? Landauer, famously skeptical of quantum computing in the early days, considered a quantum computer to be an analog device.  In his 1996 paper ``The Physical Nature of Information" he writes, ``Quantum parallelism: A return to analog computation. In quantum parallelism we do not just use 0 and 1, but all their possible coherent superpositions. This continuum range, which gives quantum parallelism its power, also gives it the problems of analog computation"~\cite{Landauer96}.  In a fundamental way, quantum systems are both analog and digital – a modern perspective of wave-particle duality -- which is what makes them so special.  Quantum states, e.g., pure states of a qubit, live in a continuum in the surface of the Bloch sphere, but the information we seek to extract is encoded in the countable (quantized) outcomes of a bit.  For a quantum computer consisting of $n$ qubits, I take the state space (the allowed inputs and outputs) to be $n$-bits -- classical information in, classical information out.  Intermediately, of course, the quantum state of the machine can exist in a superposition of classical registers, but no matter.  This is just a recipe for determining the probability of finding a given outcome; the quantum computer can be digital, but {\em nondeterministic}.  Moreover, as discussed in the classical context, all physical operations are intrinsically analog; digitization follows after coarse graining.   Dynamics occur in continuous time, governed by the time-dependent Schr\"{o}dinger equation, driven by devices that are parameterized by continuous inputs that are imperfectly calibrated.  Measurements too, even ones that are envisioned as projective onto a discrete computational basis, actually occur in continuous time with continuous signals.  Quantum digitization thus depends on our ability to coarse-grain these input and output signals, and restrict the allowed evolutions to a sequence of discrete steps from a finite alphabet. 

We may thus ask, which models of quantum computation are analog and which are digital according to the Definition, and what are the implications for computational power? In categorizing these models, we can also distinguish between universal programmable quantum computers and special purpose devices which can simulate (emulate) a specific complex quantum systems, such as condensed matter or field theory.  Emulation is often referred to as an analog quantum simulator, the quantum generalization of the classical analog integrator.  Indeed, the terms ``analog" here derives in part in that the dynamics of one quantum system is an analogy to the other.  Such a model is clearly also analog according to the Definition since it occurs in continuous time, with a continuum of Hamiltonian parameters.  Other continuous time, and thus formally analog models, include adiabatic quantum computing~\cite{Farhi00, Albash18}, quantum annealing~\cite{Kadowaki98}, and Hamiltonian quantum walks~\cite{Childs09}.

In most models of quantum computation, the input-output map we seek is the unitary propagator $U(T) = \mathcal{T}\left[ \exp\left\{-i\int_0^T H(t)\right\}\right]$ where the Hamiltonian $H(t) = \sum_i \lambda_i(t) H_i$, for some set classical control waveforms $\left\{ \lambda_i (t) \right\}$ that modulate the set of control Hamiltonians $\left\{ H_i (t) \right\}$.  Different unitary models correspond to different choices of Hamiltonian, e.g., adiabatic evolution or emulation of a target Hamiltonian. For the quantum computer to be universal, it must be ``controllable,"  meaning that for any unitary map $V$ on the Hilbert space, given the control Hamiltonians, there exist control waveforms $\{\lambda_i(t)\}$ and time $T \ge T_*$ such that $U(T)=V$~\cite{Brif10};  the minimum time $T_*$ sets the so-called ``quantum speed limit"~\cite{Deffner17}.  For a finite-dimensional Hilbert space, the set of unitary maps form a Lie group.  In that case the system is controllable if ${H_i}$ forms a set of generators of the corresponding Lie algebra.  

Two approaches to programming a universal quantum computer computer can be characterized as Hamiltonian-control~\cite{Nielsen06} (Lie algebraic) or unitary-control (Lie group). Hamiltonian-control can be achieved by through the tools of quantum optimal control, whereby the control waveforms $\{\lambda_i(t)\}$ are found numerically through optimization of a cost function~\cite{Brif10}.  Such an approach is clearly not scalable, as optimization requires numerically propagating the time-dependent Schr\"{o}dinger equation.  Nonetheless, it allows for universal programming as has been demonstrated for moderate system sizes~\cite{Lysne20}.  It is, however, clearly an analog model, with a continuum of possible control waveforms. Optimal control is also an important tool designing gates in the model described below.  Good analog control is critical for implementing reliable digital components.  

Unitary-control is known as the ``gate model" in analogy to classical logical gates for Boolean functions.  In this model, the propagator is decomposed into a sequence of unitary maps, typically acting on one or two subsystems at a time.  A universal gate set is a finite alphabet of unitary matrices $\{U_\alpha\}$ (gates) such that for any unitary map $V$ on the Hilbert space, and given an $\epsilon \ge 0$, there exists a finite sequence of gates such that $|| V - \prod_\alpha^N U_{\alpha} || \le \epsilon$.

A gate-based quantum computer is often considered synonymous with the ``digital" quantum computing. Here too, the categorization is more subtle.  A true digital model requires that the dynamics can be discretized into a finite alphabet of gates. Only in this case can error syndromes be diagnosed and corrected in a fault-tolerant manner.  However, often gate models are considered that do not satisfy this condition. For example, in quantum simulation often one employs a Trotter approximation to a desired propagator, $e^{-iT(H_A+H_b)} \approx \left(e^{-i\delta t H_A} e^{-i\delta t H_B}\right)^N$, where $\delta t = T/N$~\cite{Lloyd96}.  Each term in the Trotter expansion can be considered to be a ``gate." For variable $T$ and degree of Trotterization $N$, however, the set of gates are chosen from a {\em continuum}, and thus by definition, such a model is {\em not} digital. Indeed, Siebere {\em et al.} showed that continuous Trotterization can lead to quantum chaos for certain gate sequences with too coarse a choice of $\delta t$, which leads to a proliferation of errors, and renders gate-model implementation of a quantum simulation unreliable~\cite{Sieberer19}.   

 To be truly digital the gate model must be \textit{digitized}.  More precisely, we can consider the class of quantum computers consisting of $n$ qudits -- subsystems in $d$-dimensional Hilbert space -- corresponding to the Hilbert space $\mathcal{H} = \mathcal{C}_d^{\otimes n}$.  The state space of valid outputs is the finite alphabet of $n$ dits, $\mathbb{Z}_d^{\times n}$.  Possible dynamical maps (functions on the qudits) are in the continuous manifold describing the Lie Group $SU(d^n)$.  In the digital gate model, we choose a finite universal gate set consisting of group-generators on each single qudit, plus any entangling two-qudit unitary, taken pairwise to create a completing connect graph between all qudits.  According to the Solovay-Kitaev theorem, any $U\in SU(d)$ in the continuum of unitary matrices can be efficiently approximated by a finite sequence in the discrete gate set of universal single-qudit maps~\cite{Kitaev97, Kitaev02}. This digital gate model thus satisfies the two necessary conditions required to qualify as a digital model of computation: a finite discrete state space and a finite discrete set of transition maps that can efficiently approximate any dynamical map that has an efficient description.  As such, this model admits the possibility of quantum error correction, and fault-tolerance. Gate models that employ a continuum of possible unitary gates are {\em not digital}, they are {\em analog}, and cannot be error corrected.
 
Note, when the dimension of the Hilbert space of each subsystem is infinite, one often refers to such models as ``analog quantum computing," as the state can be envisioned as stored in an outcome that is a continuous variable (CV), such as the quadrature of a bosonic mode~\cite{Joshi20}.  In such a case one can coarse grain by encoding a qudit in an oscillator, regaining the digital structure~\cite{Gottesman01}.  Such encoding shows increasing promise,  providing error resilience that leverages the particular hardware capabilities~\cite{Campagne20, Bourassa20}.

Finally, other quantum models employ dissipation and/or measurement to achieve a desired output. For example, analog quantum simulation of equilibrium phases of strongly-correlated condensed matter using ultracold atomic gases employs cooling (dissipation of entropy) to reach complex states that that encode the order parameter~\cite{Chu18, Kantian18, Cotler19}.  The  digital gate model based on unitary dynamics can be simulated in measurement-based quantum computation (MBQC) where dynamics is driven by a sequence of single qubit measurements acting on an initially entangled ``resource state" in a basis chosen from a finite alphabet~\cite{Raussendorf01}. In an appropriate geometry, this digital MBQC model can be made fault-tolerant~\cite{Raussendorf06}.  Hybrid models that employ entangled-CV resource states, qubit encoding, and MBQC offer unique approaches to digitization, with a potential route to fault tolerance~\cite{Menicucci14}.

\subsection{Quantum Advantage in the NISQ Era}
A NISQ information processor is defined as a system that is too noisy to achieve the thresholds and scaling necessary for fault-tolerant quantum error correction, but it is sufficiently isolated from the environment and controllable that it has the potential to achieve a ``quantum advantage" over a classical information processor.  By this I mean the same as ``quantum supremacy" in the manner originally defined by Preskill:  the demonstration of a particular computation of any kind that can be done faster with a NISQ device than any current classical device~\cite{Preskill12}.  This is not a complexity theoretic definition, and of course it is a moving target that is hardware and software dependent.  Preskill laid out a number of examples where such an advantage might be seen, including optimization~\cite{Albash18, Moll17, Brando16}, quantum machine learning~\cite{Biamonte17}, and quantum simulation~\cite{Feynman82, Lloyd96, Lewenstein07, Bloch12, Blatt12, Schaetz13, Georgescu14, Johnson14, Bernien17, Monroe19}.  Currently hybrid quantum-classical models in which measurements on quantum states are inputs to a classical processor are seen as a promising arena for NISQ implementations of algorithms such as the Variational Quantum Eigensolver (VQE)~\cite{McClean16} and the Quantum Approximate Optimization Algorithm (QAOA)~\cite{Farhi14}. In addition, many workers have pursued quantum simulation of many-body physics as a near-term goal~\cite{ Lewenstein07, Bloch12, Blatt12, Schaetz13, Georgescu14, Johnson14, Bernien17, Monroe19}, both for studies of equilibrium phases of strongly-correlated matter and for studies of nonequilibrium dynamics.  Progress on NISQ processors has been steady, with improved hardware, new experimental and theoretical advances, and error mitigation strategies~\cite{Marvian17}. 

How much complexity can we harness in such NISQ systems, and what are the requirements for achieving a quantum advantage?  Is there an intermediate regime of imperfect operation, not too noisy to be classical, but not too perfect to be fault-tolerant, that allows us to transcend the power of classical computers without the full power of quantum error correction~\cite{Fujii016}?  While case-by-case analyses have been carried out, the general principles have not been laid down.  Unlike the threshold theorem for fault-tolerance, we don't know where the goalposts are and what we can achieve if we get there.  

Importantly, all NISQ processors are fundamentally analog quantum computers, by the Definition in Sec. II, regardless of whether they employ continuous time evolution or a finite sequence of gates chosen from a continuum. Coherent errors arise due to imperfect specification of the control Hamiltonian through, e.g., miscalibration, inhomogenieties, and/or quasistatic background fields, all specified by a continuum parameters. As Albash {\em et al.} have shown, analog coherent errors can have catastrophic effects on the performance of NISQ processors, e.g., in ``Ising machines" in which the solution is encoded in the ground state of an Ising model for spins connected on a graph, specified by the couplings $J_{ij}$~\cite{Albash19}.  Spin-glasses are susceptible to ``$J$-chaos," in that the ground state is hypersensitive to imperfections in the Hamiltonian parameters~\cite{MartinMayor15}.  In that case the probability that adiabatic quantum annealing will perform the desired calculation can decay exponentially with problem size for a fixed error level~\cite{Albash19}. Error mitigation schemes might reduce the detrimental effects of $J$-chaos~\cite{Pearson19}, but one should be cautious about the implication for a quantum advantage in this case, since the solutions that are amenable to error suppression may also be easier to solve classically~\cite{Preskill18}.  Similarly, in a noisy analog device we may loosen the requirement that the computation yields an exact solution.  But in that case a classical algorithm might efficiently return the same quality solution since it need only find an approximate solution to the original problem. 

These last observations raise a key question. {\em What is the relationship between the complexity of the computation to be performed and robustness of the NISQ processor to imperfections?} We expect such a relationship to exist since a quantum-supreme computation will require access to highly correlated multipartite entangled states, and it is exactly these states that are the most sensitive to decoherence. In the foundational work by Zurek on the emergence of the classical world, depending on the nature of the environment, some states can be characterized as ``pointer states," that are robust to decoherence~\cite{Zurek03}.  These states are, by this definition, classical. Are the robust states computationally simple, and what are the implications for NISQ processing?  

In the context of quantum simulation, the common lore of NISQ hardware is expressed in~\cite{Buluta09},  “…quantum simulation requires neither explicit quantum gates nor error correction, and less accuracy is needed.”  The intuition behind this statement can be understood from a thermodynamic perspective.  The solution to a quantum algorithm, like Shor’s algorithm, is often specified by an exact ``microstate” of the many-body system, i.e., the exact configuration of all spin-1/2 particles (qubits) in the computational basis.  Such a microstate is hypersensitive to perturbation where even one spin-flip is detrimental, leading to the wrong answer to the computational problem.  In contrast, a phase of matter, as studied in a quantum simulation, is characterized by an order parameter, typically specified by a few-body correlation function, e.g. a two-point correlation.   Such a ``macrostate” is consistent with a multitude of microstates and thus can be robust to small perturbations as codified in the pillars of condensed matter physics: scaling, universality, and renormalization~\cite{Stanley99}. As Preskill states, ``analog simulators are best suited for studying features that physicists call universal, properties which are relatively robust with respect to introducing small sources of error"~\cite{Preskill18}.  Macrostates are thus generally ``simple,” requiring few parameters for their description.  The challenge for condensed matter physics is to understand phenomena where such a simple description breaks down, e.g., critical phenomena and nonequilibrium dynamics.  It is in exactly these phenomena in which strongly correlated physics arises, where the system can contain highly entangled states, and which are likely to be the most fragile in the face of noise and decoherence. This begs the question: {\em what aspects of quantum simulation are simultaneously robust and computationally hard}~\cite{Hauke12}?

One way to put this question to the test is to take an adversarial approach to a quantum-supreme calculation.  Can one construct a  model for the output of the NISQ processor that can be solved efficiently classically?  Indeed, in the presence of finite temperature, control errors, and decoherence, we expect the complexity of the quantum state to be limited, and with careful choice of representation, an efficient classical solution is possible. A variety of studies have been carried out in this direction.  For example, calculating the partition function of an arbitrary ``stoquastic" Ising model is expected to be computationally hard at low temperatures~\cite{Jerrum93}. Using path integral Monte Carlo based on classical Markov chains, Crosson and Slezak showed that above a threshold temperature, independent of system size, the complexity is limited, and they constructed a fully polynomial-time approximation scheme for the partition function~\cite{Crosson20}. Similarly, in the context of optimization, Fran\c{c}a and Garcia-Patron have shown that in the presence of sufficient decoherence, there exists a nearby Gibbs thermal state from which one can  efficiently sample, and which leads to a good approximation to the minimize the cost function.  They also provide bounds that determine when this classical algorithm can beat any quantum optimizer in the presence of this noise.~\cite{Franca20}.

Another approach to designing efficient classical models is to employ tensor network representations that most efficiently capture the entanglement in the multipartite system.  The cost of classically simulating a general quantum circuit by contracting a tensor network is exponential in the treewidth of the graph induced by the circuit~\cite{Markov08}.  For restricted classes of circuits, particularly for systems that are sufficiently unentangled, efficient simulations by tensor networks are possible. In particular, for quasi-1D geometries, a matrix product representation can be used to efficiently simulate multi-qubit systems with limited entanglement that satisfy an area-law~\cite{Eisert10}.  Zhou {\em et al.} showed that one can efficiently simulate a class of quasi-1D random circuits by truncating the bond-dimension of a Matrix Product State (MPS) representation and achieve a global fidelity of $\mathcal{F} \ge 0.002$ for depth-20 1D random quantum circuits with 54 qubits~\cite{Zhou20}.  This work demonstrates that for some circuit geometries, the system only accesses a tiny fraction ($\sim 10^{-8}$) of the total Hilbert space, when the average gate infidelity is in the NISQ regime above threshold, $\epsilon \ge \epsilon_\infty\sim .01$.  

In related work, Noh {\em et al.} showed that one can employ a Matrix Product Operator (MPO) representation to efficiently simulate arbitrary 1D random circuits in the presence of depolarizing noise~\cite{Noh20}.  They demonstrate that the maximum achievable entanglement never grows larger than a constant value, which depends only on the gate error rate, not on the number of qubits nor  the depth of the circuit. Thus the maximum achievable entanglement is saturated after a certain circuit depth and consequently the required dimension of the representation does not increase exponentially with the total system size. In the 2D case, Napp {\em et al.}  showed that that for a certain family of constant depth architectures, classical simulation of typical instances is efficient if one allows for a small error, even though worst-case simulation to arbitrarily small error is intractable~\cite{Napp20}. As they explain, ``The intuitive reason for this is that the simulation of 2D shallow random circuits can be reduced to the simulation of a form of an effective 1D dynamics which includes random local unitaries and weak measurements. The measurements then cause the 1D process to generate much less entanglement than it could in the worst case, making efficient simulation possible."  In addition, they speculate that in the presence of noise, their algorithms may be able to simulate larger depths and qudit dimensions than in the noiseless case.
 
These results quantitatively demonstrate what is intuitively true. Systems that are more ``classical" can have an efficient representation.  This classicality can occur because the unitary dynamics does not generate substantial multipartite entanglement, and/or because the system is open, and decoherence reduces the nonclassical features.  These efficient representations provide the framework for efficient simulations.  Importantly, it is critical to consider what one is simulating. One may not be interested in simulating the exact quantum state. In a quantum simulation, generally one is simulating an expectation value, typically a one- or two-point correlation function that represents an order parameter.  In that situation an efficient simulation must accurately capture the few-body reduced density operator. For weakly correlated systems, this sets an even a lower bar for classical simulation. In such systems we expect that we can severely truncate the higher order correlations with little effect on the low order moments.  In contrast, for strongly-correlated matter, we expect that such truncation will not be possible.  But, these higher correlations will be the most sensitive to decoherence, and thus the output will be hypersensitive to imperfections.  The reliability of NISQ quantum simulation is thus most in doubt for exactly those problems that are hardest to simulate classically, but robust for those cases that have efficient representations. 

An open quantum system that is not digitized and error corrected will never scale; nonclassical features will saturate and become independent of system size.  This is not to say that a NISQ system cannot achieve ``quantum supremacy," in the manner Preskill originally defined it~\cite{Preskill12}.  For sufficiently high fidelity operation, for some tasks a quantum computer could exceed the power of the current most powerful classical computer.  Have we reached this threshold?  Building on foundational studies of the complexity of sampling~\cite{Aaronson13, Bouland19}, Google has claimed quantum supremacy in the context of sampling from a probability distribution as generated from a 2D random quantum circuit~\cite{Arute19}.  No current classical computer can efficiently generate samples from the distribution generated by the ideal noiseless circuit.  If it could, a major breakdown of the complexity of the polynomial hierarchy would occur. Clearly, the noisy Google circuit is not producing samples from the ideal distribution, but a nearby distribution.  Though the variational distance to the nearby distribution may be small, this does not imply that it is impossible to efficiently draw samples from the noisy distribution with an appropriately constructed algorithm. 

How low must the noise be and what are the restrictions on the noise model such that efficient sampling from the output distribution is prohibited?  While some computational complexity arguments indicate that it implausible that a classical computer can spoof Google's linear cross-entropy benching test~\cite{Aaronson19}, the challenge remains open.  The results of Noh {\em et al.}~\cite{Noh20} indicate that efficient simulations are possible for open quantum systems in 1D and those of Napp {\em et al.}~\cite{Napp20} show a possible extension to arbitrary 2D random circuits.  If we restrict to a closed quantum system, Zlokapa {\em et al.}~\cite{Zlokapa20} employed the Schr\"{o}dinger-Feynman algorithm to show that more efficient classical simulations of 2D random circuits are possible allowing for finite fidelity. For current fidelities in the Sycamore processor, a quantum advantage for cross-entropy benchmarking limits to around 300 qubits.  An attempt to challenge Google's supremacy result with classically efficient models will help us to understand where the frontier lies.  

In the absence of error correction, quantum coherence will decay with time and depth of the quantum circuit.  Thus, one of the most pressing questions is to show that NISQ devices can solve hard problems with low depth circuits (short time compared to fidelity decay time).  Can we unambiguously challenge the Extended Church-Turing Thesis by demonstrating a clear separation between classical and quantum computation for some problem with some constraint of resources?  The Google quantum supremacy experiment employs strongly believed arguments based on the complexity hierarchy to study this in the context of sampling the output of a low-depth 2D random circuit~\cite{Arute19}.  Another approach taken by Bravyi and coworkers is to study how one can achieve a quantum advantage with shallow constant depth circuits when compared to classical algorithms that have equal depth~\cite{Bravyi20}.  While this approach is not equivalent to ``quantum supremacy" in the sense of demonstrating a calculation that cannot be done efficiently on current classical computers, it does demonstrate a rigorous separation between quantum and classical complexity.   In particular, they show that certain nonlocal games can be solved with such shallow 1D constant-depth circuits whereas a classical algorithm would require the depth to grow logarithmically in the number of qubits.  Such multiplayer nonlocal games achieve a quantum advantage through a combination of nonlocality and {\em contextuality}, the fact that the measurement outcome of an observable depends on the ``context" of other observables measured at the same time.  This power has been extended by Daniel and Miyake to a more general class of states with symmetry-protected topological order (1D SPTO), which shows how contextuality can lead to unconditional computational separation for sufficiently large string order parameters~\cite{Daniel20}.  

Another critical issue is to understand how errors propagate.  For a NISQ processor that is designed to solve a complex problem, we don't know how to propagate errors, because the output probability is too complex to calculate.  So currently, we don't know if we can trust the output of NISQ devices, and when we can, what the implications are for the complexity of the solution. One way to explore this is to consider narrow but {\em deep} quantum circuits.  Jessen and coworkers have developed a Small, Highly Accurate, Quantum (SHAQ) simulator based on universal control of a 16-dimensional Hilbert space (isomorphic to 4 qubits) associated with the spin of individual cesium atoms, in which they perform $\sim 100$ arbitrary SU(16) gates, each with fidelity $>0.99$~\cite{Lysne20}.  These experiments demonstrate how certain observables can be robust to imperfections, while the overall state fidelity decays.  Working with this platform Poggi {\em et al.} established rigorous criteria to show how imperfect devices are able to reproduce the dynamics of macroscopic observables accurately, while the relative error in the expectation value of microscopic observables is much larger on average~\cite{Poggi20}.  The next challenge is to extend these ideas to rigorously establish how this macroscopic/microscopic divide is related to a robustness/complexity trade-off.  

\section{Summary  and Outlook}
Quantum Information Science grew out of decades of fundamental curiosity-driven research.  How is quantum mechanics different from classical statistical mechanics?  Are there local hidden variables that explain random measurement outcomes?  How does the impossibility superluminal communication constrain quantum information? What is the meaning of negative probability?  Are measurement outcomes contextual?  How do we understand the emergence of the classical world from a fundamentally quantum one? Answers to these questions did not change the formalism of quantum mechanics, but they did change how we thought about it.  Simultaneously, the development of information science forced us to consider the implications that physics has on computation and other information processing tasks and vice versa.  Turing, Church, and Shannon abstracted information from the devices that processed it, but thermodynamics and quantum mechanics forced us to understand the role of physics in information.  And the reverse is true.  Information science has taught us to understand statistical physics and quantum mechanics.  No physicists would have believed that we could prevent spontaneous decay by error correction. Basic research has laid out the foundations upon which the second quantum revolution rests.

Since the establishment of the foundations, concepts from QIS have led to new understandings in other fields, notably in many-body physics and field theory.  Entanglement plays an essential role in explaining critical phenomena and phases of matter~\cite{Zeng19}.  The power of the density matrix renormalization group (DMRG)~\cite{White92} derives from the efficient representation of entanglement encoded in matrix product states~\cite{Verstraete08}.  The extension  to general tensor networks represents one of the most powerful tools for describing complexity in quantum systems~\cite{Orus14}.  This extends to new descriptions of spacetime itself and the famous AdS/CFT correspondence in quantum gravity and string theory, which has the structure of a quantum error correcting code~\cite{Almheiri15}.

As the second quantum revolution transitions from science to technology, a strong foundation of curiosity-based basic research is essential for success.  A desire to deliver near-term NISQ applications should not deter critical analysis to determine whether this is even possible.  Fundamental questions remain unanswered.  How much complexity can we truly harness from a noisy quantum system?  What is the relationship between robustness of a NISQ device, and the complexity to obtain that output?  What is the computational complexity of analog quantum devices?  What role do different quantum resources play in attaining a quantum advantage: entanglement, negativity, noncontextuality?  A deeper understanding of these questions will create a richer framework for developing useful applications. 

Explorations of these questions will also impact basic physics.  How much quantum complexity does Nature harness? Natural systems, such as condensed matter materials are imperfect, subject to noise an decoherence. Do we really need a quantum supreme simulator to explain high temperature superconductivity, or does the existence of this robust macroscopic order parameter indicate that in fact an efficient (approximate) classical description awaits us. Perhaps only an engineered fault-tolerant quantum computer can lift the observable effects of highly entangled complex states to the macroscopic world, or perhaps there are special cases, such as topologically protected states that are exceptions to the rule~\cite{Hasan10}.  Understanding how we can can harness quantum complexity for technology will guide our ability to understand the natural world and vice versa.  If Nature has some tricks up its sleeve, we should borrow them.

The next stage of basic research will take unexpected turns, as recent history has already shown. The seminal work of Haroche and collaborators probed how ``cat-states" (superpositions of coherent states) decohere~\cite{Haroche96}.  The Ramsey interferometric technique they developed to observe the "birth and death of photon" in a cavity~\cite{Haroche07} is now a method for syndrome extraction on an error-correcting ``cat-code"~\cite{Mirrahimi14}  which was the first to demonstrate ``break-even" in correcting native errors~\cite{Ofek16}. This example demonstrates how curiosity-based research leads to applications never imagined in the initial study.  It also points to a key component for progress in QIS -- interdisciplinarity.  A deeper collaboration between physicists, chemists, computer scientists, electrical engineers, mathematician, theorists, modelers, experimentalists, and device engineers is essential for us to bridge the gap between fundamental science and technology.  Harnessing the second quantum revolution will require us to move out of our comfort zone, learn new things, and synthesize the perspectives of others. The second quantum revolution is accelerating and with eyes wide open to the opportunities and real challenges, the revolution will deliver.

\section*{Acknowledgments}
I gratefully acknowledge my colleagues, 
Tameem Albash and Pablo Poggi for their insights,  critical reading, and feedback on this manuscript.  I also thank numerous other colleagues and students with whom I have debated over the years regarding these concepts including Poul Jessen, Carl Caves, Dave Bacon, Ken Brown, Akimasa Miyake, Anupam Mitra, Bob Keating, Gopi Muraleedharan, Sayonee Ray, Elizabeth Crosson, Andrew Landahl, and Robin Blume-Kohout. This work was supported by the NSF Focused Research Hub for Theoretical Physics, CQuIC, Grant PHY-1630114.

\bibliography{apssamp.bib}

\end{document}